# All-fiber hydrogen sensor based on stimulated Raman gain spectroscopy with a 1550 nm hollow-core fiber

Fan Yang*[a,b], Wei Jin[a,b]

[a]Department of Electrical Engineering, The Hong Kong Polytechnic University, Hung Hom, Kowloon, Hong Kong, China; [b] Photonic Sensors Research Center, The Hong Kong Polytechnic University Shenzhen Research Institute, Shenzhen, 518057, China

## ABSTRACT

We report a highly sensitive all-fiber hydrogen sensor based on continuous-wave stimulated Raman gain spectroscopy with a hollow-core photonic crystal fiber operating around 1550 nm. A pump-probe configuration is used, in which the frequency difference between the pump and the probe lasers is tuned to the $S_0(0)$ transition of para-hydrogen with a Raman shift of 354 cm$^{-1}$. Preliminary experiments demonstrate a detection limit down to 17 ppm with a 250 s averaging time, and further improvement is possible. The all-fiber configuration operating in the telecommunication wavelength band would enable cost-effective and compact sensors for high sensitivity and high-resolution trace analysis.

## 1. INTRODUCTION

Raman spectroscopy is a versatile tool for analyzing molecules and materials. It is commonly used to provide a fingerprint by which molecules can be identified. When the laser light interacts with molecular vibrations or rotations, light energy would be shifted up or down and the shift in energy gives information about the vibrational or rotational modes of the molecules.

Both spontaneous Raman scattering and stimulated Raman scattering (SRS) have been studied for trace-gas detection. However, Raman scattering is typically very weak and the discrimination of spontaneous Raman scattering against background such as Rayleigh scattering is poor especially for pure rotational Raman bands that have small Raman shifts. The spectral resolution of the spontaneous Raman sensor is limited by the spectrometer used for analysis [1].

In SRS, two laser beams with different frequencies (i.e., $\omega_p$ for pump and $\omega_s$ for Stokes) are used. The frequency difference, $\Delta\omega = \omega_p - \omega_s$, is tuned to a particular Raman transition and consequently, the intensity of the Stokes beam experiences a gain (stimulated Raman gain, SRG) and the intensity of the pump beam experiences a loss (stimulated Raman loss, SRL). The gain or loss is proportional to the pump power, the light-gas interaction length and gas concentration. Compared with spontaneous Raman scattering, SRG has very high spectral resolution limited by the linewidth of the lasers used. SRS has been studied for a wealth of applications, including Raman gas laser [3], high resolution gas spectroscopy [4] and high sensitivity label-free microscopic imaging [5].

Recently, there have been several reports on the use of hollow-core photonic crystal fibers (HC-PCFs) to enhance the sensitivity of Raman-based (spontaneous and stimulated) gas sensors. Improvement of detection sensitivity by two to three orders of magnitude over the free-space systems has been demonstrated [6-9], and detection limit of 100-1000 ppm for a number of gases (e.g., $H_2$, $N_2$, $O_2$, $CO_2$) at ambient conditions has been achieved. Under high pressure (e.g., 20 bars) conditions, gas detection down to several ppm has been reported with a meter-long HC-PCF and a high power (2W) pump beam at 532 nm [10]. However, these systems operate in visible [6, 8, 10], 800 nm [7] or 1000 nm band [9], and bulk optical components were used except the sensing HC-PCFs.

In this paper, we report an all-fiber gas sensor system based on SRG spectroscopy with a commercial HC-1550-06 HC-PCF. The HC-PCF has a transmission window from 1500 to 1650 nm, broad enough to cover the rotational transitions of some important gases such as $N_2$, $O_2$, $CO_2$. We present here the results of a hydrogen sensor based on the $S_0(0)$ rotational Raman transition of para-hydrogen.

*fpolyyang@polyu.edu.hk; phone 852 9711-4390; fax 852 2330-1544;

## 2. EXPERIMENTAL SETUP

The experimental setup for gas detection with SRG spectroscopy is shown in Fig. 1(a). A distributed feedback (DFB) semiconductor laser is amplified by an Er-doped fiber amplifier (EDFA) and used as the pump source and it wavelength is modulated at 50 kHz. The wavelength of the DFB is tuned to around 1532.244 nm. An external-cavity diode laser is used as the probe beam and its wavelength tuned to 1620.22 nm. The frequency (wavelength) difference matches the $S_0(0)$ rotational Raman transition of para-hydrogen with Raman shift $\Omega = 354.36$ cm$^{-1}$[11]. The probe beam is detected by the photo-detector (PD) and demodulated by a lock-in amplifier.

The gas cell is constructed with a 15-m-long HC-1550-06 fiber with core diameter of ~11 μm, and the SEM image of the HC-PCF is shown in the inset of Fig. 1(a). The output end of the sensing HC-PCF is fusion spliced to a single-mode fiber (SMF) with low fusion current to prevent the collapse of air-holes while the input end is butt coupled to an input SMF with a mechanical splicer. Gas is pressurized into the HC-PCF through the gap between HC-PCF and input SMF. The normalized transmission spectrum of the 15-m-long HC-PCF gas cell with SMF pigtails is shown in Fig. 1(b). The total loss from the input SMF to the output SMF is ~6 dB at the pump and probe wavelengths.

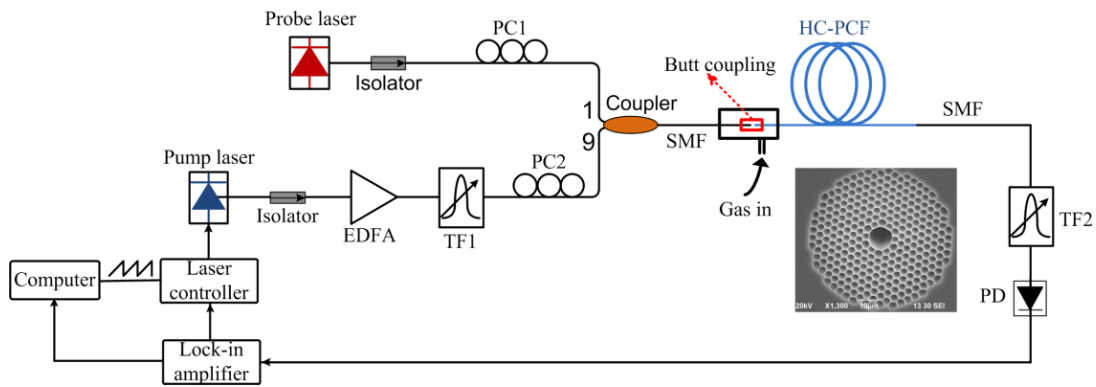

(a)

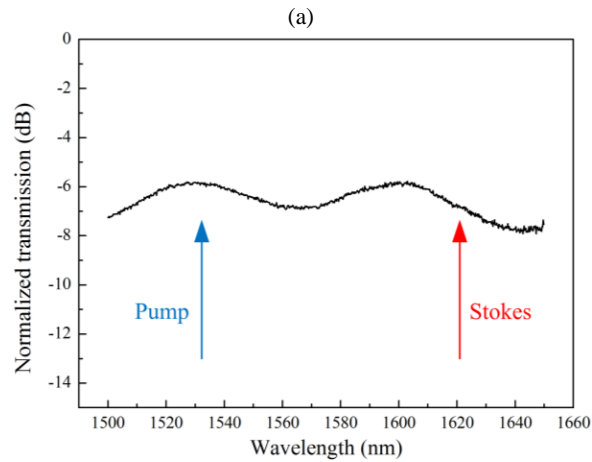

(b)

Figure 1. (a) Experimental setup for gas detection with 15-m-long HC-PCF. TF1 is used to minimize the EDFA's amplified spontaneous emission (ASE) noise and TF2 is used to filter out the residual pump. PC1 and PC2 are used to maximize the Raman signal. EDFA, erbium-doped fiber amplifier; PD, photodetector; TF, tunable filter; PC, polarization controller; SMF, single-mode fiber. The inset shows the scanning electronic microscopy image of the HC-1550-06 fiber. (b) Normalized transmission spectrum of the gas cell made with 15-m-long HC-1550-06 fiber.

## 3. RESULTS

Figure 2(a) shows the second-harmonic lock-in outputs for different pump power levels delivered to the HC-PCF when the pump wavelength was tuned across 1532.244 nm and the probe wavelength was fixed to 1620.22 nm. The HC-PCF was filled with high purity hydrogen (Scientific Grade, hydrogen concentration > 99.999% from ARKONIC) to a uniform pressure of 1.6 bar. The power of the probe beam at the photo-detector is ~50 μW. For the pump power of 36.6 mW, the peak amplitude of the second-harmonic signal is ~1.60 mV. The peak amplitude signal and the standard deviation of the noise as functions of pump power level when the pump is tuned to 1532.211 nm are shown in Fig. 2(b). The signal amplitude increases linearly with pump power while the noise level shows very little change. The standard deviation of the noise over 3 minute duration is 0.226 μV, which is not much larger than the noise level (0.217 μV) when the pump is off and is ~8 times the noise level (0.028 μV) when both the pump and probe are off. The lower detection limit in terms of noise equivalent concentration (NEC) is estimated to be 141 ppm for 1 s lock-in time constant.

We also conducted an Allan variance analysis based on the measured second-harmonic signal over a one-hour period when the pump wavelength was tuned to 1532.211 nm. The result is shown in Fig. 3. For an averaging time of ~150 seconds, the noise is 0.04 μV and the corresponding NEC is 25 ppm. For an averaging time of ~250 seconds, the noise is 0.028 μV and the corresponding NEC is 17 ppm.

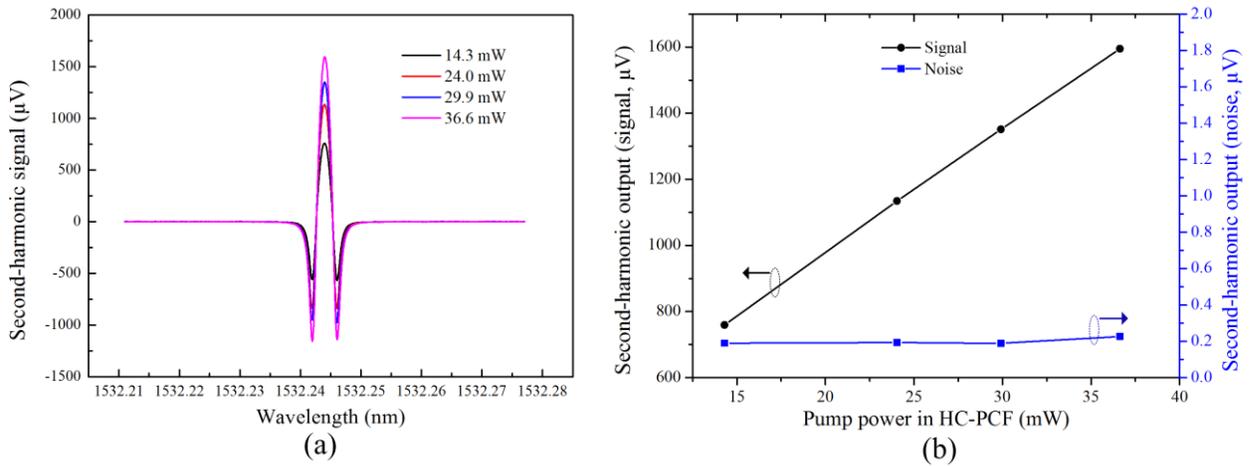

Figure 2. Experimental results for 15-m-long HC-PCF gas cell. (a) Second-harmonic lock-in output (signal) when the pump laser is tuned across 1532.244 nm. (b) Second-harmonic signal and the standard deviation of the noise when the pump wavelength is fixed to 1532.211 nm as functions of pump power level in HC-PCF.

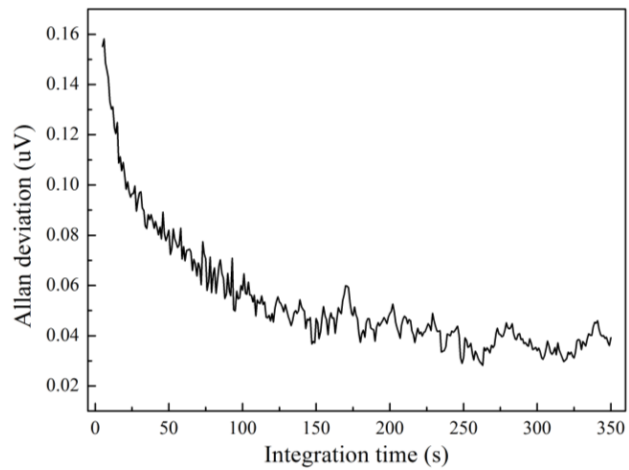

Figure 3. Allan deviation plot of the sensor system.

## 4. DISCUSSION

We calculated the shot noise limit of the current system and found it is ~8.8 time better than the experimental result. This indicates that there is room for further improvement, since we have made no effort in optimizing the system configuration and the parameters to minimize the various noise factors.

As shown in Fig. 2(b), the signal is linearly proportional to the pump power level while the noise shows little change. This agrees with the theory [8] that Raman gain is proportional to the pump power as well as gas concentration. In our preliminary experiments, the maximum pump power into the HC-PCF is 36.6 mW, and we believe that ppm level hydrogen detection is achievable by using a higher power pump source in combination with an optimized detection system.

HC-PCF-based SRG spectroscopy can readily be applied to detect other types of gases. The HC-1550-06 fiber we used here has a transmission window from 1500 to 1650 nm, which covers the rotational Raman shifts of a range of gases such as $N_2$, $O_2$, $CO_2$ and could then be used to detect these gases.

## 5. SUMMARY

An all-fiber hydrogen gas detection system using low power CW lasers operating in the telecommunication band is reported. The NEC is 141 ppm for 1 s lock-in time constant, and goes down to 17 ppm for 250 s averaging time. The system could achieve ppm level hydrogen detection if a higher level of pump power is used, which would have potential applications in for example transformer condition monitoring in electrical power industry.

## ACKNOWLEDGMENTS

This work was supported by National Natural Science Foundation of China (NSFC) through Grant No. 61535004 and Grant No. 61290313 and the Hong Kong SAR government through GRF grant PolyU 152603/16E.

## REFERENCES

[1] Latka, I, Dochow, S., Krafft, C., Dietzek, B. and Popp, J., "Fiber optic probes for linear and nonlinear Raman applications – current trends and future development," Laser Photon. Rev. 7(5), 698-731 (2013).
[2] Maier, M., "Applications of stimulated Raman scattering," Appl. Phys. 11(3), 209-231 (1976).
[3] Meng, L. S., Roos, P. A. and Carlsten, J. L., "Continuous-wave rotational Raman laser in $H_2$," Opt. Lett. 27(14), 1226-1228 (2002).
[4] Owyoung, A. and Jones E. D., "Stimulated Raman spectroscopy using low-power cw lasers," Opt. Lett. 1(5), 152-154 (1977).
[5] Freudiger, C. W., et al., "Label-free biomedical imaging with high sensitivity by stimulated Raman scattering microscopy," Science 322(5909), 1857-1861 (2008).
[6] Buric, M. P., Chen, K. P., Falk, J. and Wooddruff, S. D., "Enhanced spontaneous Raman scattering and gas composition analysis using a photonic crystal fibre," Appl. Opt. 47(23), 4255-4261 (2008).
[7] Yang, X., Chang, A. S. P., Chen, B., Gu, C., Bond, T. C., "High sensitivity gas sensing by Raman spectroscopy in photonic crystal fiber," Sensors and Actuators B 176, 64-68 (2013).
[8] Doménech, J. L. and Cueto, M., "Sensitivity enhancement in high resolution stimulated Raman spectroscopy of gases with hollow-core photonic crystal fibres," Opt. Lett. 38(20), 4074-4077 (2013).
[9] Westergaard, P. G., Lassen, M. and Peterson, J. C., "Differential high resolution stimulated CW Raman spectroscopy of hydrogen in hollow-core fiber," Opt. Express 23(2), 16320-16328 (2015).
[10] Hanf, S., et al., "Fast and highly sensitive fiber-enhanced Raman spectroscopic monitoring of molecular $H_2$ and $CH_4$ for point-of-care diagnosis of malabsorption disorders in exhaled human breath," Anal. Chem. 87, 982-988 (2015).
[11] Carlsten, F. J. L. and Wenzel, R. G., "Stimulated rotational Raman scattering in $CO_2$-pumped para-$H_2$," IEEE J. Quantum Electron. QE-19(9), 1407-1413 (1983).